\documentclass[prd,showpacs,preprintnumbers,amsmath,amssymb,11pt]{revtex4}

\usepackage{graphics}
\usepackage[utf8]{inputenc}
\usepackage{epsfig}
\usepackage{subfigure}
\usepackage{dcolumn}
\usepackage{bm}
\usepackage{color}

\begin{document}

\title{Hunting for the $X_b$ via  Radiative Decays}

\author{Gang Li$^{1}$
and
Wei Wang$^{2}$
}

\affiliation{1) Department of Physics, Qufu Normal University, Qufu
273165, People's Republic of China}

\affiliation{2) Helmholtz-Institut f\"ur Strahlen- und Kernphysik
and Bethe Center for Theoretical Physics, Universit\"at Bonn,
D-53115 Bonn, Germany}

\begin{abstract}
In this paper, we study   radiative  decays of  $X_b$, the counterpart of the famous $X(3872)$ in the bottomonium-sector as a candidate for meson-meson molecule, into the $\gamma \Upsilon(nS)$ ($n=1$, $2$, $3$). Since it is likely that  the $X_b$ is  below the $B\bar B^*$ threshold and the mass difference between the neutral and charged  bottom meson is small compared to the binding energy of the $X_b$, the isospin violating decay mode $X_b\to \Upsilon (nS)\pi^+\pi^-$ would be  greatly suppressed. This  will  promote  the importance of the radiative decays.  We use  the effective Lagrangian based on the heavy quark symmetry to explore the rescattering mechanism and  calculate  the partial widths. Our results show that the partial widths into $\gamma \Upsilon(nS)$ are about $1$ keV,  and thus the branching fractions may be sizeable, considering the fact the total width may also be smaller than a few MeV like the $X(3872)$.  These radiative decay modes are of great importance in the experimental search for the $X_b$ particularly  at hadron collider. An observation of the $X_b$ will provide a deeper   insight into the exotic hadron spectroscopy and is helpful to  unravel  the nature of the states connected  by the heavy quark symmetry.

\end{abstract}

\date{\today}







\maketitle

\section{Introduction}
\label{sec:introduction}

In the past decades, there has been great progress in hadron spectroscopy thanks to the unprecedented data sample accumulated by the B factories and hadron-hadron colliders. A number of charmonium-like and bottomonium-like states have been discovered  on these experimental facilities so far but  not all of them can be placed in  the ordinary $\bar qq$ (for reviews, see Refs.~\cite{Brambilla:2010cs,Godfrey:2008nc,Drenska:2010kg,Bodwin:2013nua}).

The $X(3872)$ is the first and perhaps the most renowned exotic candidate. It  was first discovered  in 2003  by  Belle  in the $B^+\to K^++ J/\psi \pi^+\pi^-$ final state~\cite{Choi:2003ue} and subsequently  confirmed
by the BaBar Collaboration~\cite{Aubert:2004ns}.
Complementary observation is also found in   proton-proton/antiproton
collisions at the Tevatron~\cite{Abazov:2004kp,Aaltonen:2009vj} and
LHC~\cite{Chatrchyan:2013cld,Aaij:2013zoa}.   Though the existence is well established, the  nature of the $X(3872)$ is still ambiguous due to  a few  peculiar properties. First, compared to typical hadronic widths the total width is  tiny. Only an upper bound has been measured experimentally:
$\Gamma<1.2$ MeV~\cite{Beringer:1900zz}. The mass lies closely to the
$D^0\overline D^{*0}$ threshold,
$M_{X(3872)}-M_{D^0}-M_{D^{*0} }=(-0.12\pm0.24)$~MeV~\cite{TheBABAR:2013dja}, which leads  to
speculations that  the $X(3872)$  is presumably  a meson-meson molecular state~\cite{Tornqvist:2004qy,Hanhart:2007yq}.

These peculiar  features have stimulated considerable  research interest in investigating the production and decays of  the $X(3872)$ towards understanding  its nature. A very  important aspect  involves the discrimination of a compact multiquark configuration and a loosely bound hadronic molecule configuration.
In this viewpoint,  it would  be also valuable  to look for the  analogue in the bottom sector, referred to as $X_b$  following the notation suggested in Ref.~\cite{Hou:2006it}, as states related by heavy quark symmetry may have  universal behaviours.
Since the $X_b$ is  expected to be very heavy and its $J^{PC}$ of  is $1^{++}$, it is less likely  for a  direct discovery  at the current electron-positron collision facilities, though the Super KEKB may provide an opportunity in   $\Upsilon(5S,6S)$ radiative decays~\cite{Aushev:2010bq}.

In Ref.~\cite{GMW}, the production  of the $X_b$ at the LHC and the Tevatron has been investigated, along the same line with   the  studies  on the search for exotic states at hadron colliders~\cite{Bignamini:2009sk,Artoisenet:2009wk,
Artoisenet:2010uu,Esposito:2013ada,Ali:2011qi,Ali:2013xba,Guo:2013ufa}.
It is shown  that the production rates at the LHC and the Tevatron are sizeable~\cite{GMW}.  On the other hand, the  search for the $X_b$ also depends on   reconstructing the $X_b$, which motivates us to   study the  $X_b$ decays. Since this meson is expected to be far below threshold, the isospin violating decay mode for instance $X_b\to \Upsilon\pi^+\pi^-$ is highly  suppressed, and this may explain the escape of $X_b$ in the recent CMS search~\cite{Chatrchyan:2013mea}.  As a consequence,   radiative decays of the $X_b$ will be of high priority,  on which we will focus in this paper.  As we will show in the following, these modes have sizeable decay widths.

To calculate the radiative decays, we study  the intermediate meson loop contributions, which have been one of the
important nonperturbative transition mechanisms in various transitions,
and their impact on the  heavy quarkonium transitions, also referred to as
coupled-channel effects, has been noticed for a long
time~\cite{Lipkin:1986bi,Lipkin:1988tg,Moxhay:1988ri}. The intermediate meson loops mechanism has been applied to study the production and decays of ordinary and  exotic states~\cite{Guo:2009wr,Guo:2010ak,Wang:2013cya,Liu:2013vfa,Guo:2013zbw,Wang:2013hga,Cleven:2013sq,Chen:2011pv,Li:2012as,Li:2013yla,Voloshin:2013ez,Voloshin:2011qa,Bondar:2011ev,oai:arXiv.org:1002.2712,Chen:2011pu,Chen:2012yr,Chen:2013coa,Chen:2013bha} and B decays~\cite{Du:1998ss,Chen:2000ih,Liu:2007qs,Lu:2005mx,Colangelo:2003sa,Liu:2008tv,Cheng:2004ru,Colangelo:2002mj}, and a global agreement with experimental data is found. Thus this approach may be an effective approach to handle the $X_b$ radiative decays.

The paper is organized as follows. In Sec.~\ref{sec:formula}, we
will introduce the formalism used in this work.  Based on this framework,  numerical results are presented in Sec.~\ref{sec:results} and the summary will be given in
Sec.~\ref{sec:summary}.

\section{Radiative decays}
\label{sec:formula}

\begin{figure}[hbt]
\begin{center}
\includegraphics[width=0.8\textwidth]{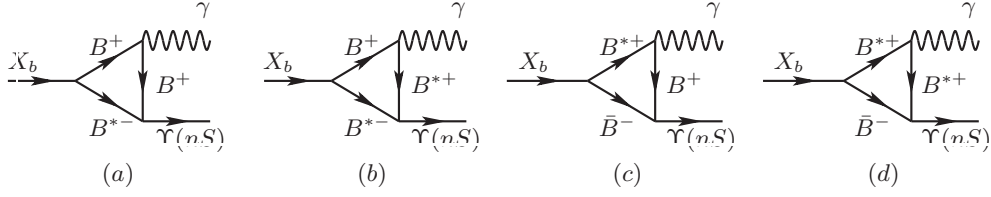}
\caption{Feynman  diagrams for  the radiative decays $X_b \to \gamma \Upsilon(nS)$ with  the $B{\bar B}^*$ as the intermediate states.} \label{fig:loops}
\end{center}
\end{figure}

The  calculation of contributions from the
meson loops requests  the leading order effective Lagrangian. Based on the heavy quark symmetry,  we employ the relevant
effective Lagrangian for the $\Upsilon(nS)$~\cite{Colangelo:2003sa,Casalbuoni:1996pg}
\begin{eqnarray}
\mathcal{L}_{\Upsilon(nS) B^{(*)} B^{(*)}} &=&
ig_{\Upsilon BB} \Upsilon_{\mu} (\partial^\mu B \bar{B}- B
\partial^\mu \bar{B})-g_{\Upsilon B^* B} \varepsilon^{\mu \nu
\alpha \beta}
\partial_{\mu} \Upsilon_{\nu} (\partial_{\alpha} B^*_{\beta} \bar{B}
 + B \partial_{\alpha}
\bar{B}^*_{\beta})\nonumber\\
&&-ig_{\Upsilon B^* B^*} \big\{
\Upsilon^\mu (\partial_{\mu} B^{* \nu} \bar{B}^*_{\nu}
-B^{* \nu} \partial_{\mu}
\bar{B}^*_{\nu})+ (\partial_{\mu} \Upsilon_{\nu} B^{* \nu} -\Upsilon_{\nu}
\partial_{\mu} B^{* \nu}) \bar{B}^{* \mu}  \nonumber\\
&& +
B^{* \mu}(\Upsilon^\nu \partial_{\mu} \bar{B}^*_{\nu} -
\partial_{\mu} \Upsilon^\nu \bar{B}^*_{\nu})\big\}, \label{eq:h1}
\end{eqnarray}
where
${{B}^{(*)}}=\left(B^{(*)+},B^{(*)0}\right)$ and
${\bar B^{(*)T}}=\left(B^{(*)-},\bar{B}^{(*)0}\right)$ correspond to the
bottom meson isodoublets. { $\epsilon^{\mu\nu\alpha\beta}$ is the anti-symmetric Levi-Civita tensor and $\epsilon^{0123}= -1$. Due to the heavy quark symmetry, the following relationships of the couplings are valid~\cite{Casalbuoni:1996pg,Colangelo:2003sa}
\begin{eqnarray}
g_{\Upsilon(nS) BB} = 2g_n \sqrt{m_{\Upsilon(nS)}} m_B \ ,
\quad g_{\Upsilon(nS) B^* B} = \frac {g_{\Upsilon(nS) BB}} {\sqrt{m_B m_{B^*}}} \ ,
\quad g_{\Upsilon(nS) B^* B^*} = g_{\Upsilon(nS) B^* B}  \sqrt{\frac {m_{B^*}} {m_B}} m_{B^*},
\end{eqnarray}
where $g_n = \sqrt{m_{\Upsilon(nS)}}/(2m_B f_{\Upsilon(nS)})$;
$m_{\Upsilon(nS)}$ and $f_{\Upsilon(nS)}$ denote the mass and decay constant of
$\Upsilon(nS)$, respectively. The decay constant $f_{\Upsilon(nS)}$ can
be extracted from the $\Upsilon(nS)\to e^+e^-$:
\begin{eqnarray}
\Gamma(\Upsilon(nS) \to e^+e^-) = \frac {4\pi\alpha^2} {27} \frac {f_{\Upsilon(nS)}^2} {m_{\Upsilon(nS)}},
\end{eqnarray}
where $\alpha = 1/137$ is the electromagnetic fine-structure constant. Using the masses and
leptonic decay widths of  the $\Upsilon(nS)$ states: $\Gamma(\Upsilon(1S)
\to e^+e^-) =1.340 \pm 0.018$ keV, $\Gamma(\Upsilon(2S) \to e^+e^-)
=0.612 \pm 0.011$ keV, $\Gamma(\Upsilon(3S) \to e^+e^-) =0.443 \pm
0.008$ keV~\cite{Beringer:1900zz}, one can  obtain $f_{\Upsilon(1S)} =
715.2 $ {\rm MeV}, $f_{\Upsilon(2S)} = 497.5 $ {\rm MeV}, and
$f_{\Upsilon(3S)} = 430.2 $ {\rm MeV}.}

We consider the iso-scalar $X_b$ as a $S$-wave molecular state with the positive charge parity given by the superposition of $B^0 {\bar B}^{*0}+c.c$ and $B^- {\bar B}^{*+}+c.c$ hadronic configurations as
\begin{eqnarray}
|X_b\rangle= \frac {1} {2} [  (|B^0{\bar B}^{*0}\rangle - |B^{*0} {\bar B}^0\rangle) +   (| B^+ B^{*-}\rangle - | B^- B^{*+}\rangle ) ].
\end{eqnarray}
The coupling of $X_b$ to the bottomed meson is based on  the effective  Lagrangian
\begin{eqnarray}
{\cal L} = \frac {1} {2} X_{b\mu}^{\dagger} [x_1(B^{*0\mu} {\bar B}^0 - B^{0} {\bar B}^{*0\mu})+x_2(B^{*+\mu} B^- - B^+ B^{*-\mu})] + h.c.,
\end{eqnarray}
{  where $x_i$ denotes the  coupling constant. }

For a bound  state  below an $S$-wave two-hadron threshold, the effective coupling of this state to the two-body channel is related to the probability of finding the two-hadron component in the physical wave function of the bound states and the binding energy, $E_{X_b}=m_B+m_{B^*}-m_{X_b}$~\cite{Weinberg:1965zz, Baru:2003qq,Guo:2013zbw}
\begin{eqnarray}\label{eq:coupling-Xb}
x_i^2 \equiv 16\pi (m_B+ m_{B^*})^2 c_i^2 \sqrt{\frac {2E_{X_b}}{\mu}} ,
\end{eqnarray}
where $c_i=1/{\sqrt 2}$, $\mu=m_Bm_{B^*}/(m_B+m_{B^*})$ is the reduced mass.

The magnetic coupling of the photon to heavy bottom  meson   is
described by the Lagrangian~\cite{Hu:2005gf,Amundson:1992yp}
\begin{eqnarray}
{\cal L}_\gamma = \frac {e\beta Q_{ab}} {2} F^{\mu\nu} {\rm Tr}[H_b^\dagger \sigma_{\mu\nu} H_a  ] + \frac {e Q^\prime} {2m_{Q}} F^{\mu\nu} {\rm Tr}[H_a^\dagger H_a \sigma_{\mu\nu}],
\end{eqnarray}
{  with
\begin{eqnarray}
H&=&\left( \frac{1+ \rlap{/}{v} }{2} \right)
[\mathcal{B}^{*\mu}
\gamma_\mu -\mathcal{B}\gamma_5],
\end{eqnarray}}
where $Q= {\rm diag}\{2/3, -1/3, -1/3\}$ is the light quark charge
matrix, $\beta$ is an unknown parameter  and $Q^\prime$ is the heavy quark
electric charge (in units of $e$).  {  In the nonrelativistic constituent quark model $\beta\simeq 3.0$ GeV$^{-1}$, which has been adopted  in the study of radiative $D^*$ decays~\cite{Amundson:1992yp}.  Note heavy quark symmetry ensures that $\beta$ is the same in the $b$ and $c$ systems, so we take the same value as Ref.~\cite{Amundson:1992yp}.} The first term is the magnetic
moment coupling of the light quarks, while the second one  is the
magnetic moment coupling of the heavy quark and hence is
suppressed by $1/m_Q$.

The decay amplitudes for the transitions in
Fig.~\ref{fig:loops} can be expressed in a generic  form in the effective Lagrangian approach as follows,
\begin{eqnarray}
M_{fi}=\int \frac {d^4 q_2} {(2\pi)^4} \sum_{B^* \ \mbox{pol.}}
\frac {V_1V_2V_3} {a_1 a_2 a_3}{\cal F}(m_2,q_2^2)
\end{eqnarray}
where $V_i$ and $a_i = q_i^2-m_i^2 \ (i=1,2,3)$ are the vertex
functions and the denominators of the intermediate meson
propagators. For example, in Fig.~\ref{fig:loops} (a), $V_i \
(i=1,2,3)$ are the vertex functions for the initial $X_b$, final
bottominum and photon, respectively. $a_i \
(i=1,2,3)$  are the denominators for the intermediate $B^+$,
$B^{*-}$ and $B^+$ propagators, respectively. In addition, we introduce a dipole form
factor,
\begin{eqnarray}\label{ELA-form-factor}
{\cal F}(m_{2}, q_2^2) \equiv \left(\frac
{\Lambda^2-m_{2}^2} {\Lambda^2-q_2^2}\right)^2,
\end{eqnarray}
where $\Lambda\equiv m_2+\alpha\Lambda_{\rm QCD}$ and the QCD energy
scale $\Lambda_{\rm QCD} = 220$ MeV. This form factor is supposed to
compensate the off-shell effects arising from
the intermediate exchanged particle and the non-local effects of the
vertex functions~\cite{Li:1996yn,Locher:1993cc,Li:1996cj}, and phenomenological studies have suggested $\alpha\sim 2$. The explicit expression of the transition amplitudes can be found in Appendix (A.6) in Ref.~\cite{Zhao:2013jza}, where radiative decays of charmonium are studied extensively based on the effective Lagrangian approach.

\section{Numerical Results}
\label{sec:results}

The existence of the $X_b$ was predicted in both the tetraquark
model~\cite{Ali:2009pi} and hadronic molecular
calculations~\cite{Tornqvist:1993ng,Guo:2013sya,Karliner:2013dqa}. The mass of
the lowest-lying $1^{++}$ $\bar b \bar q bq$ tetraquark was predicted to be
10504~MeV in Ref.~\cite{Ali:2009pi}, while the mass of the $B\bar B^*$
molecule based on the mass of the $X(3872)$ is a few tens of MeV
higher~\cite{Guo:2013sya,Karliner:2013dqa}. In Ref.~\cite{Guo:2013sya}, the
mass was predicted to be $(10580^{+9}_{-8})$~MeV, corresponding to a binding
energy of $(24^{+8}_{-9})$~MeV. These studies have provided a range for the binding energy, for which in the following we will choose a few illustrative values: $E_{X_b} =(1, 2, 5, 20)$ MeV.

\begin{figure}[ht]
\centering
\includegraphics[width=0.45\textwidth]{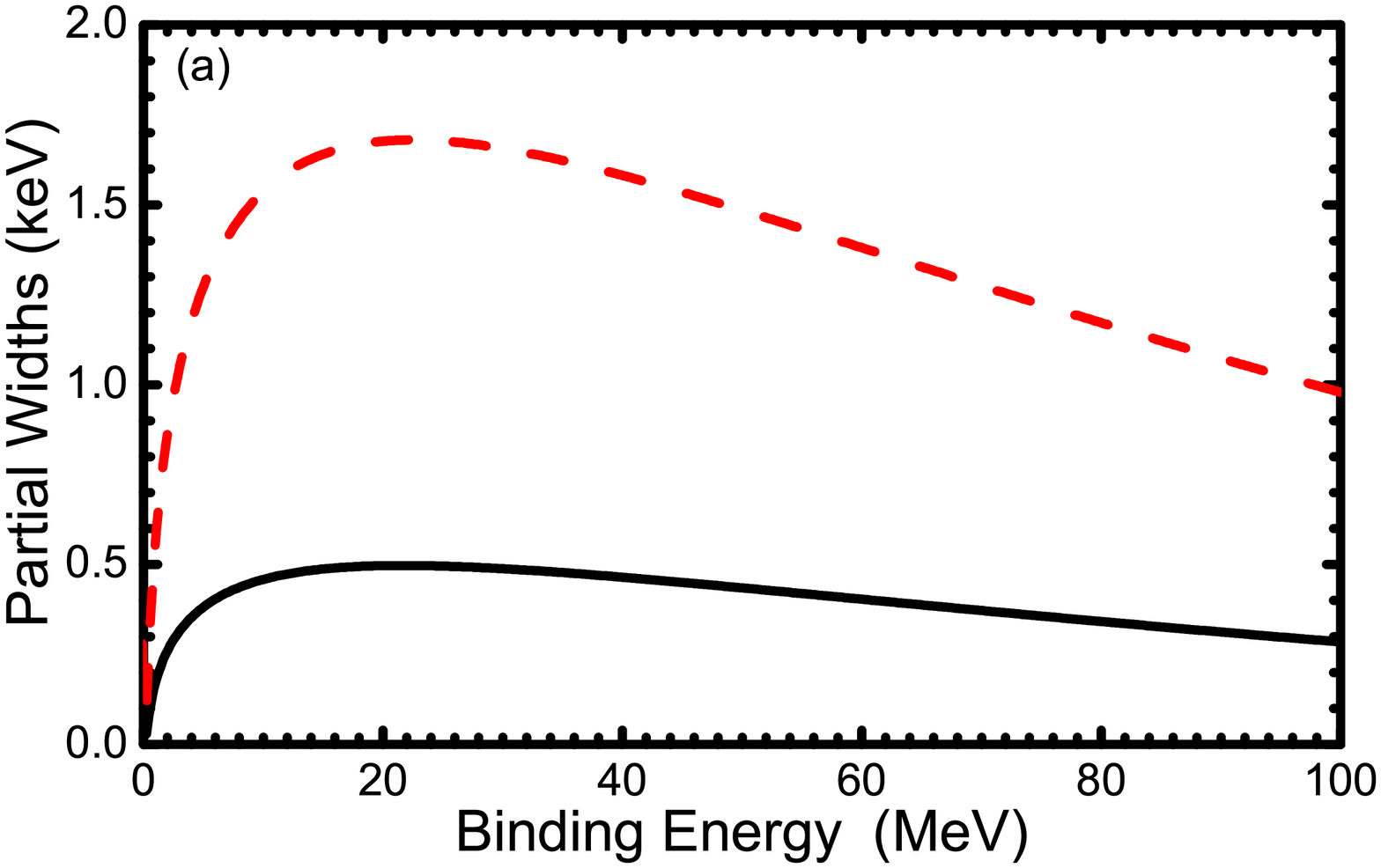}
\includegraphics[width=0.45\textwidth]{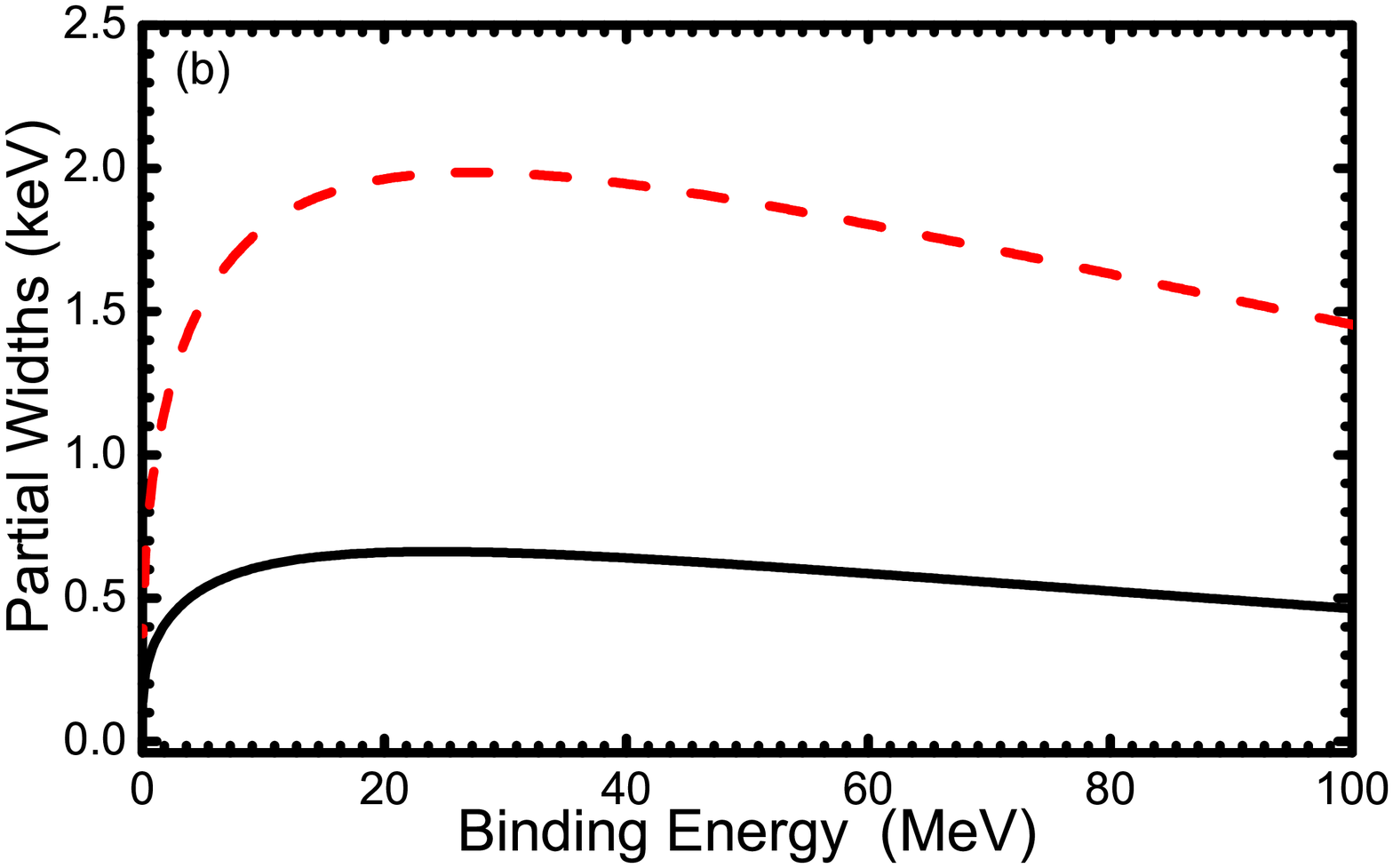}\\
\includegraphics[width=0.45\textwidth]{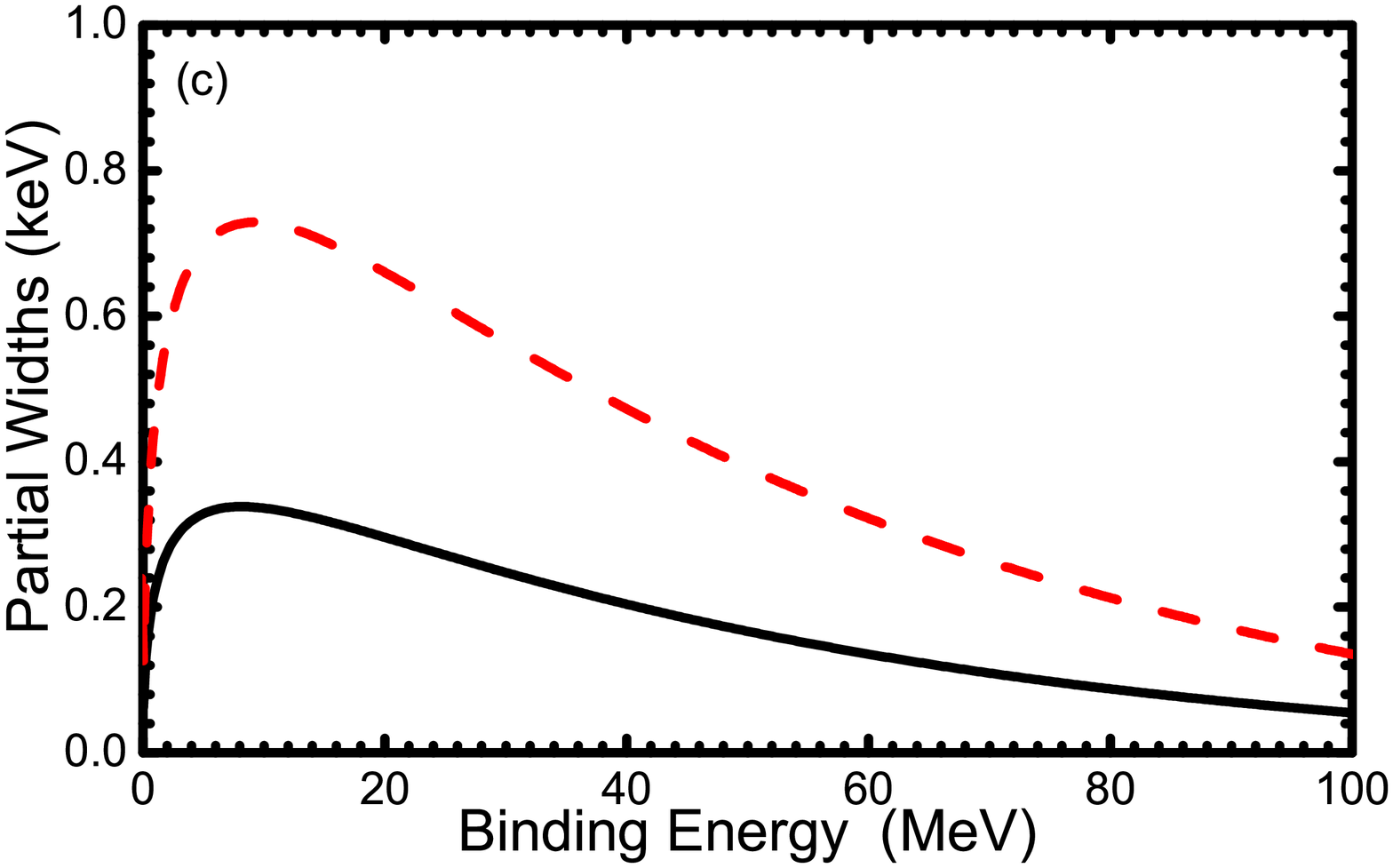}
\caption{ The dependence of partial widths of
$X_b \to \gamma\Upsilon(1S)$ on the $E_{X_b}$ with $\alpha=2.0$ (solid lines) and $\alpha=3.0$ (dashed lines), respectively. Panels (b) and (c) corresponds to the ones in the $X_b \to \gamma\Upsilon(2S)$  and $3S$, respectively. }\label{fig:WidthOnEXb}
\end{figure}

\begin{figure}[ht]
\centering
\includegraphics[width=0.45\textwidth]{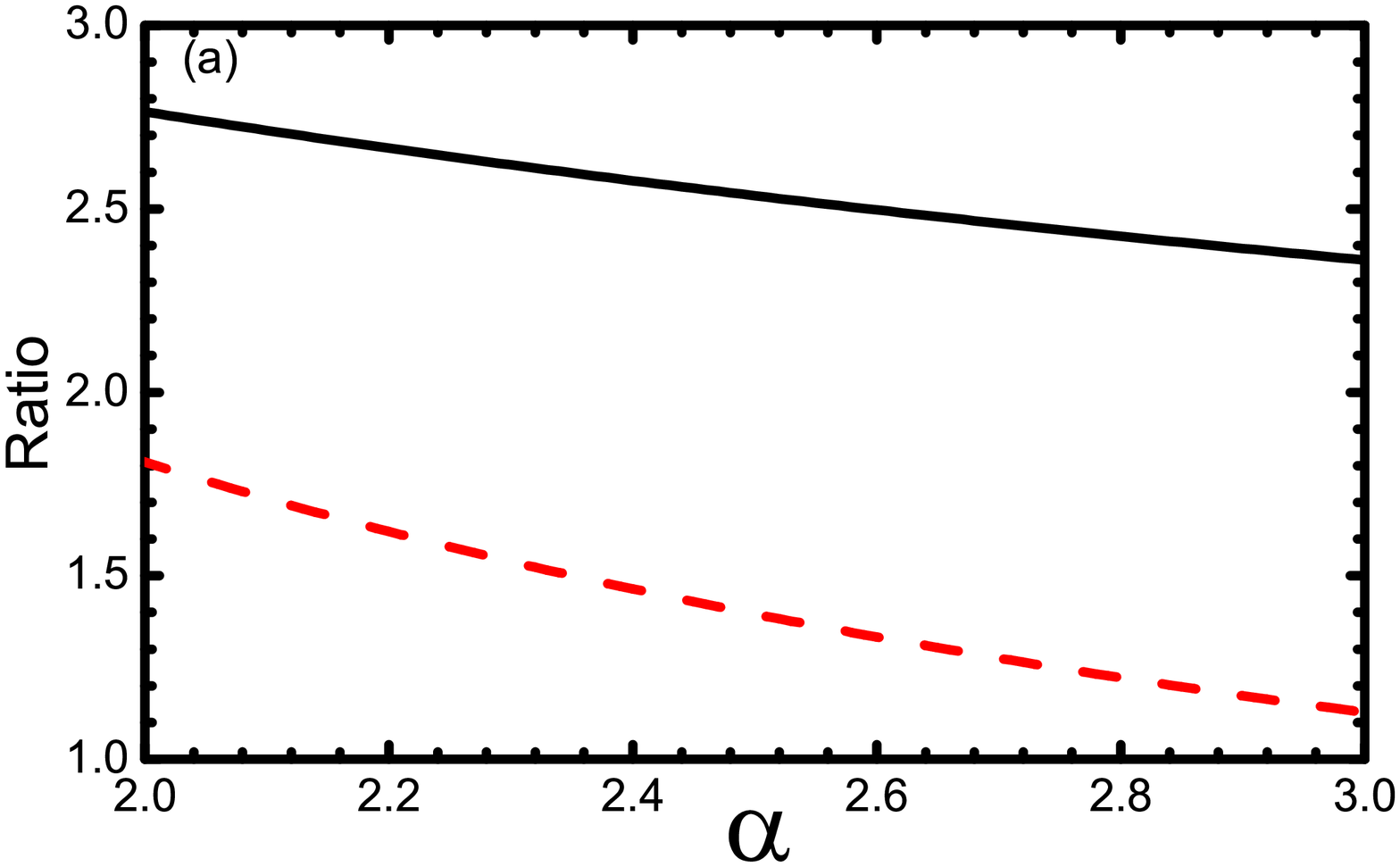}
\includegraphics[width=0.45\textwidth]{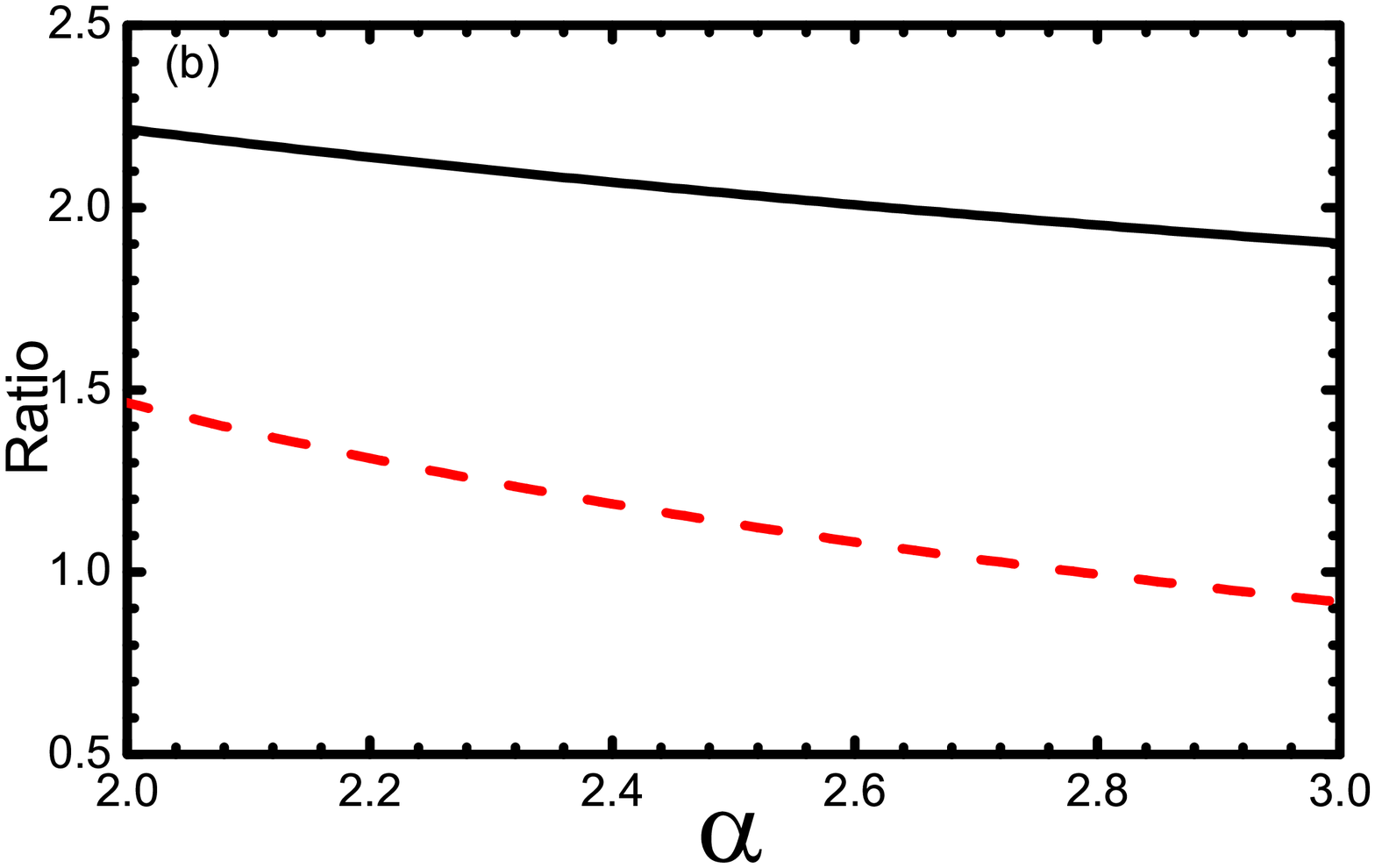} \\
\includegraphics[width=0.45\textwidth]{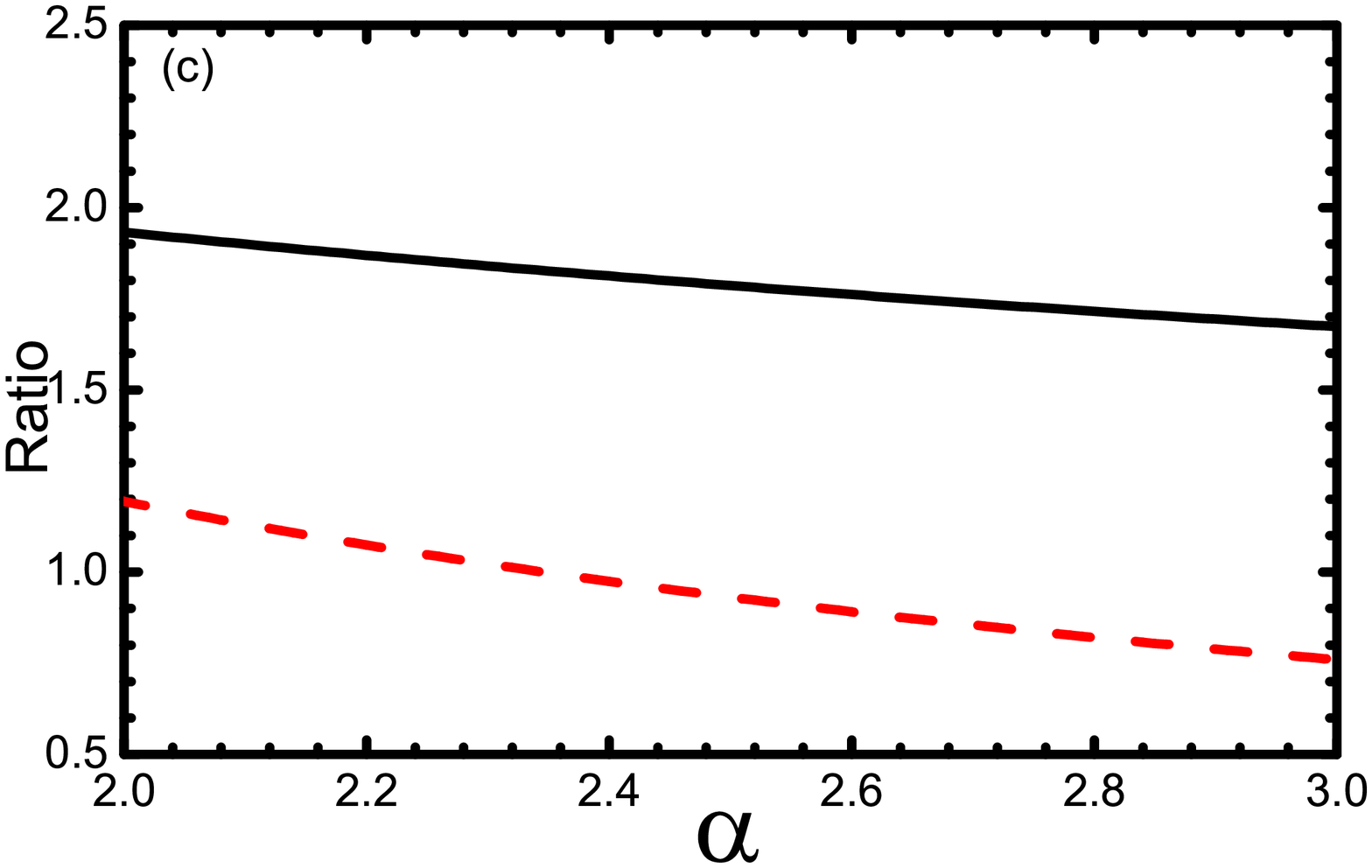}
\includegraphics[width=0.45\textwidth]{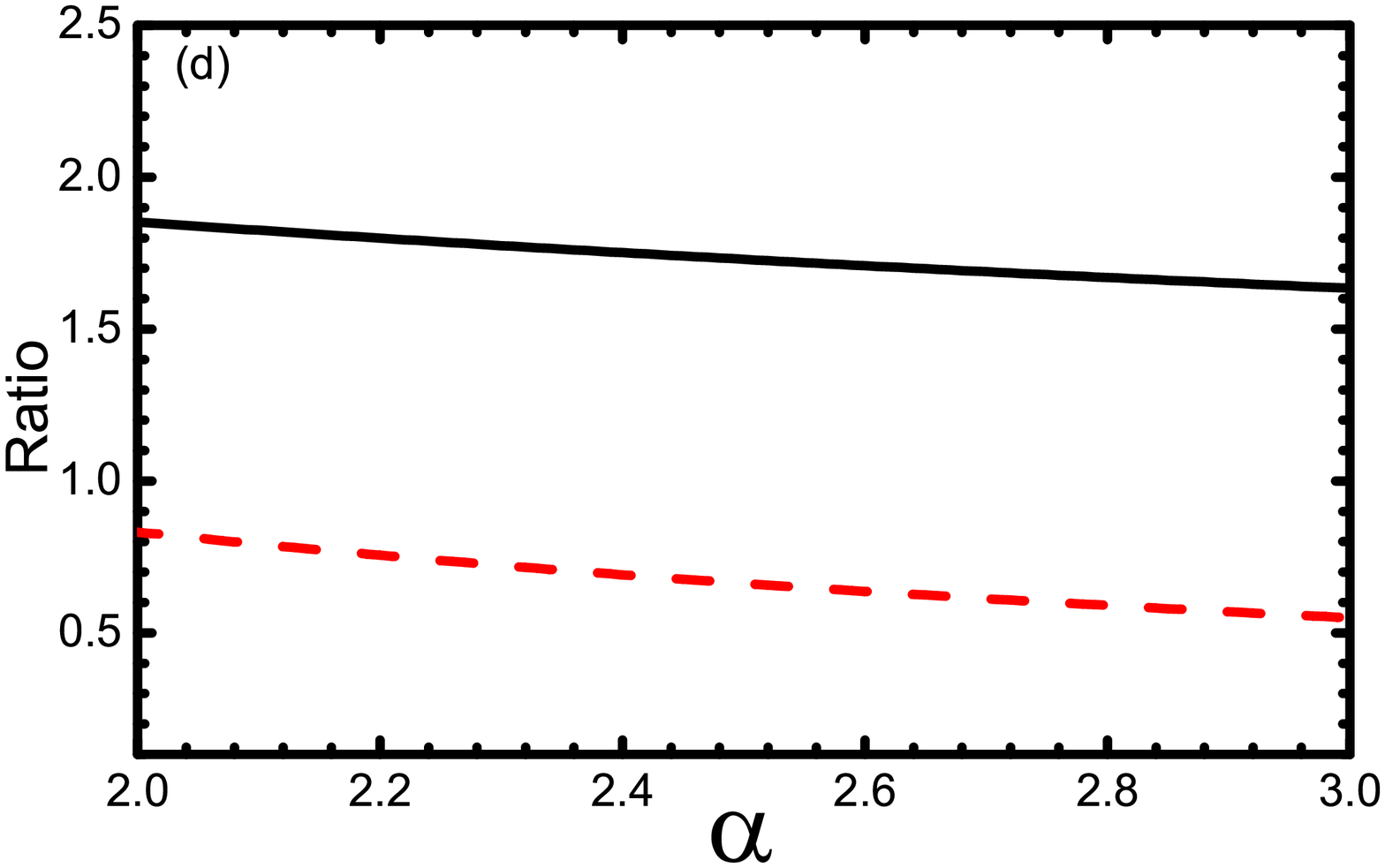} \\
\caption{ (a) The $\alpha$-dependence of the ratios of $R_1$ (solid line), and $R_2$ (dashed line) defined in Eq.~(\ref{eq:ratio}) with $E_{X_b}=1$ MeV. (b), (c), and (d) corresponds to $E_{X_b}=2$ MeV, $5$ MeV, and $20$ MeV, respectively.}\label{fig:3}
\end{figure}

\begin{figure}[ht]
\centering
\includegraphics[width=0.45\textwidth]{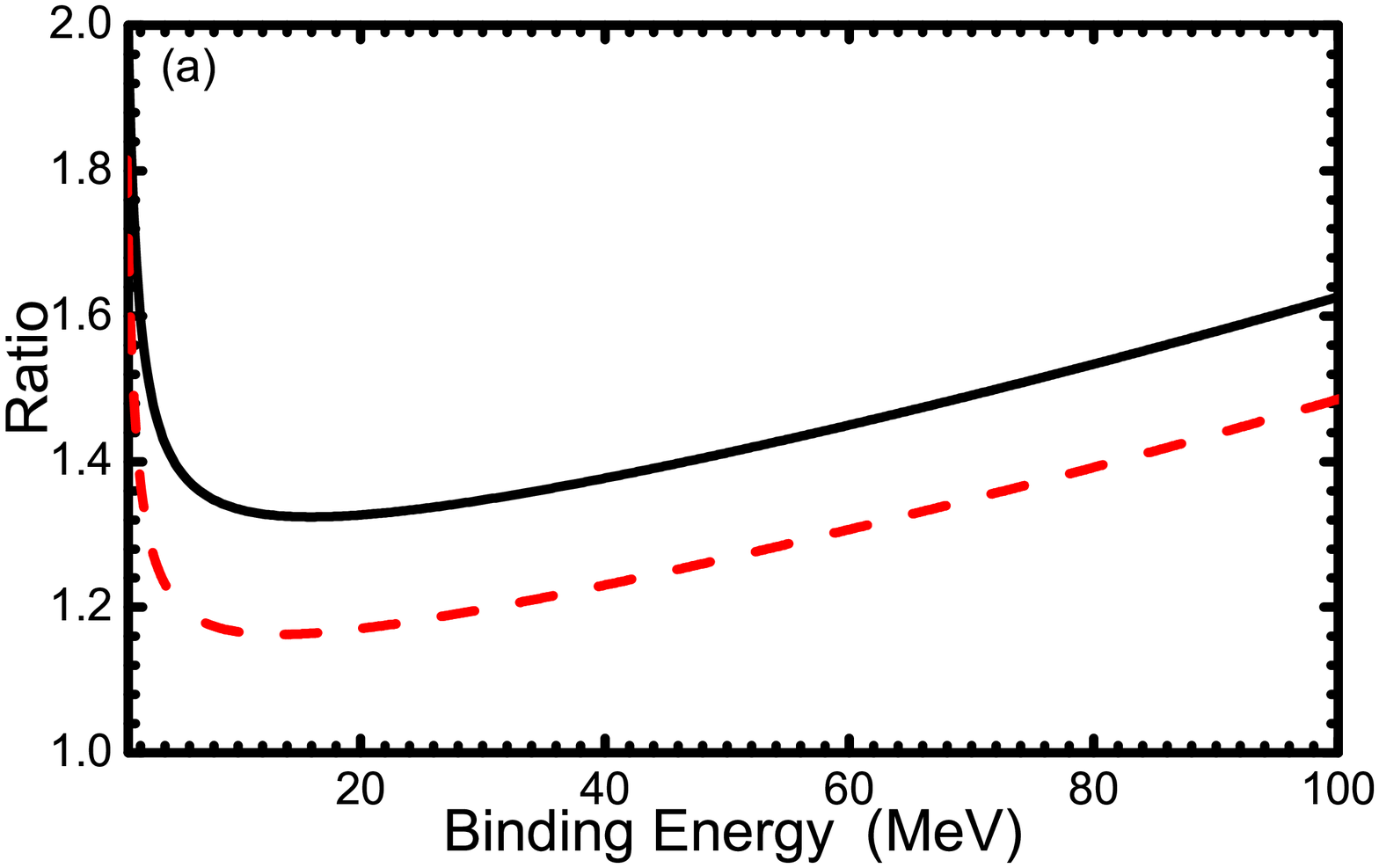}
\includegraphics[width=0.45\textwidth]{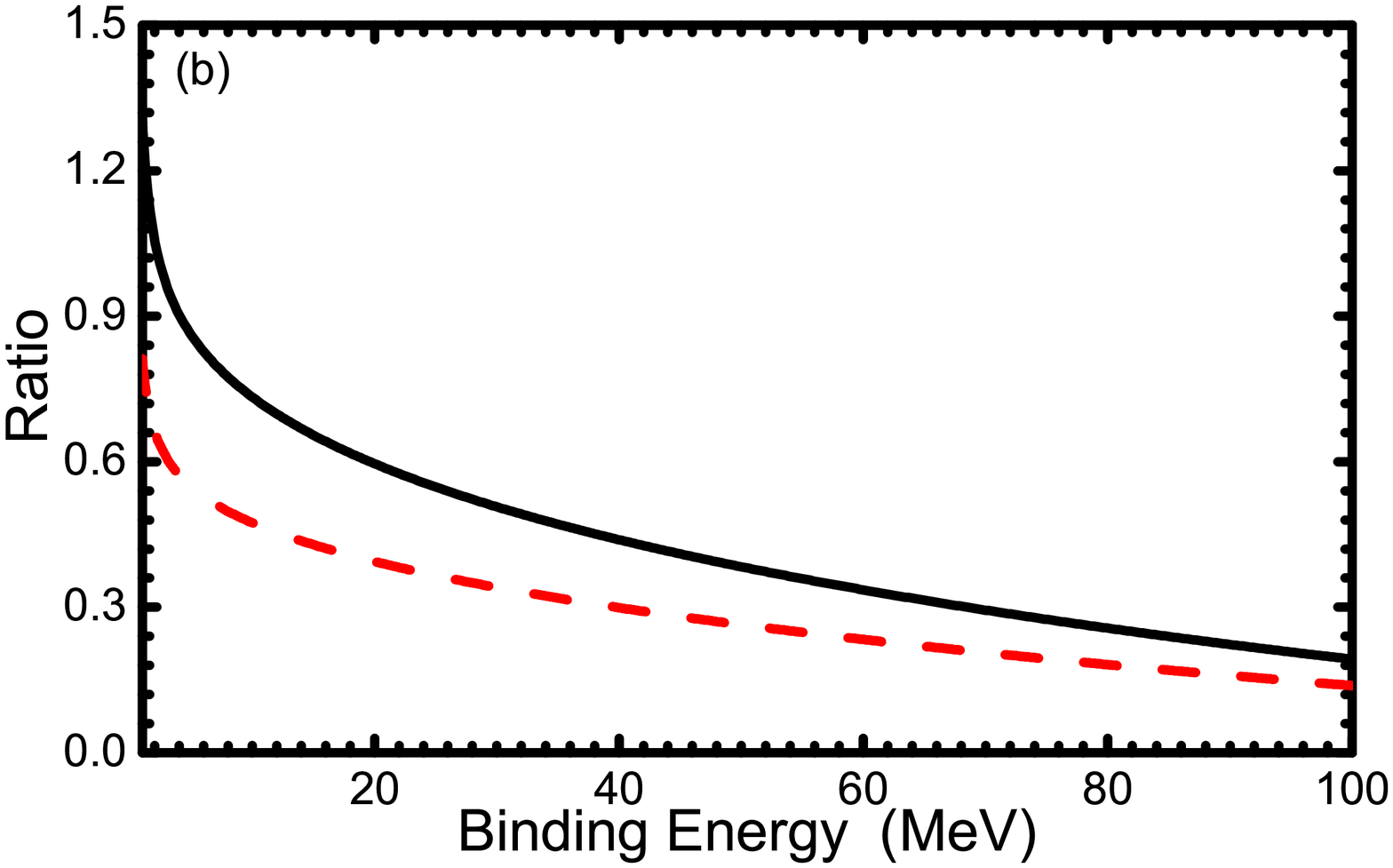} \\
\caption{ (a) The ratio $R_1$ defined in Eq.~(\ref{eq:ratio}) in terms of the $E_{X_b}$ with $\alpha=2.0$ (solid line) and $\alpha=3.0$ (dashed line). (b) The same notation with (a) except for $R_2$ defined in Eq.~(\ref{eq:ratio}).}\label{fig:ratio_mass_R}
\end{figure}

\begin{table}[htb]
\begin{center}
\caption{Predicted partial widths (in unit of keV) of the $X_b$ decays.  The  parameter in the form factor is chosen as $\alpha =2.0$ and $\alpha =3.0$. }\label{tab:results}
\begin{tabular}{ccccccc}
\hline
& \multicolumn{3}{c} {$\alpha=2.0$} & \multicolumn{3}{c} {$\alpha=3.0$} \\ \hline
& $\gamma\Upsilon(1S)$ & $\gamma\Upsilon(2S)$ & $\gamma\Upsilon(3S)$ & $\gamma\Upsilon(1S)$ & $\gamma\Upsilon(2S)$ & $\gamma\Upsilon(3S)$ \\ \hline
$E_{X_b} = 1$ MeV  & 0.12  & 0.34 & 0.22  & 0.41 & 0.96 & 0.46 \\ \hline
$E_{X_b} = 2$ MeV  & 0.19 &0.42 & 0.28& 0.62 &1.18 & 0.57 \\ \hline
$E_{X_b} = 5$ MeV  & 0.28 &0.53& 0.33 & 0.92& 1.53 & 0.70 \\ \hline
$E_{X_b} = 20$ MeV & 0.36 & 0.66 & 0.30 & 1.20 & 1.96& 0.66 \\ \hline
\end{tabular}
\end{center}
\end{table}

Choosing two values for the cutoff parameter $\alpha$, we have predicted the partial decay widths and the  numerical results are collected in Table~\ref{tab:results}. From this table, we can see that the widths for the $X_{b}$ radiative decays  are about $1$ keV. It is noteworthy to recall that the upper bound for the $\Gamma(X(3872))$ is $1.2$ MeV~~\cite{Beringer:1900zz}. If the $X_{b}$ were similarly narrow, our results would indicate a sizeable branching fractions, at least $10^{-3}$, for these radiative decay modes.

In Fig.~\ref{fig:WidthOnEXb}, we present the partial widths for the  $X_b\to \gamma \Upsilon(1S)$ (panel a), $\gamma\Upsilon(2S)$ (panel b), and $\gamma\Upsilon(3S)$ (panel c) in terms of the  $E_{X_b}$ with $\alpha=2.0$ (solid lines) and $3.0$ (dashed lines), respectively. The uncertainties caused by the cutoff parameter  indicate  our limited knowledge on the applicability of the effective Lagrangian.
However fortunately the $\alpha$ dependence of the partial widths are not drastically sensitive, which indicates a reasonable cutoff of the ultraviolet contributions by the empirical form factors. In this figure,  there exists  an evident enhancement structure around $E_{X_b}=20$ MeV resulting from the cusp effect. As can be seen from this figure, this enhancement structure is independent of the cutoff parameter $\alpha$.

It would be interesting to further clarify the uncertainties arising from the introduction of the form factors by studying  the ratios between different partial decay widths. We define the following ratios
\begin{eqnarray}
R_1= \frac {\Gamma(X_b\to \gamma\Upsilon(2S))} {\Gamma(X_b\to \gamma\Upsilon(1S))}, \quad R_2= \frac {\Gamma(X_b\to \gamma\Upsilon(3S))} {\Gamma(X_b\to \gamma\Upsilon(1S))}, \label{eq:ratio}
\end{eqnarray}
which are plotted in Fig.~\ref{fig:3} for the dependence on the cutoff parameter and Fig.~\ref{fig:ratio_mass_R} for the dependence on binding energy. Since the first coupling vertices are the same for those decay channels when taking the ratio, so the ratio only reflects the open threshold effects through the intermediate bottomed meson loops. The ratios are less sensitive to the cutoff parameter, which is a consequence of the fact that the involved loops are the same. As can be seen from this figure, when the cutoff parameter $\alpha$ increases, the ratios decrease. These predictions can be tested by the experimental measurements in future.

\section{Summary}
\label{sec:summary}

Our understanding of hadron spectroscopy will  be greatly improved by  studies of  exotic  states  that may defy the conventional models of $q\bar q$ meson spectroscopy, and accordingly   great progress  has been made in the past decades.
One of the most important aspects in the study of exotics  is the
discrimination of a compact multiquark configuration and a loosely bound
hadronic molecule. Such task requests a large  amount of   efforts on both experimental and theoretical sides in future.

In this work,  we have  investigated  the radiative  decays of the $X_b$, the counterpart of the famous $X(3872)$ in the bottomonium-sector as a candidate for meson-meson molecule, into the $\gamma \Upsilon(nS)$.   Since this state may be far below the $B\bar B^*$ threshold, the isospin violating   decay mode $X_b\to \Upsilon\pi^+\pi^-$ would be highly suppressed, and stimulate the importance of the radiative decays.  We have made used of  the effective Lagrangian based on the heavy quark symmetry, and explore the rescattering mechanism. Our results have shown that the partial widths for the $X_b\to \gamma \Upsilon(nS)$ are about $1$ keV, and thus the branching fractions may  be sizeable, taking into account  the fact the total width may also be smaller than a few MeV like $X(3872)$.
This study of radiative decays and the previous work on production rates in hadron-hadron collisions have indicated  a promising  prospect to find the $X_b$ at hadron collider in particular the LHC, and we suggest our experimental colleagues to perform an analysis. Such attempt  will likely lead to the discovery of the $X_b$ and thus enrich the exotics garden in the heavy quarknoium sector.


\section*{Acknowledgements}
\label{sec:acknowledgements}

The authors are very grateful to Feng-Kun Guo, Xiao-Hai Liu, Qian Wang,  and Qiang Zhao  for useful discussions. W.W. thanks Ulf-G. Mei{\ss}ner and Feng-Kun Guo for the collaboration of Ref.~\cite{GMW}.
This work is supported in part by the National Natural Science Foundation of China (Grant No. 11275113),
the China Postdoctoral Science Foundation (Grant No. 2013M530461), and  the DFG and the NSFC through funds provided to
the Sino-German CRC 110 ``Symmetries and the Emergence of Structure in QCD''.


\begin{thebibliography}{99}



\bibitem{Brambilla:2010cs}
  N.~Brambilla, S.~Eidelman, B.~K.~Heltsley, R.~Vogt, G.~T.~Bodwin, E.~Eichten, A.~D.~Frawley and A.~B.~Meyer {\it et al.},
  Eur.\ Phys.\ J.\ C {\bf 71}, 1534 (2011)
  [arXiv:1010.5827 [hep-ph]].

\bibitem{Godfrey:2008nc}
  S.~Godfrey and S.~L.~Olsen,
  Ann.\ Rev.\ Nucl.\ Part.\ Sci.\  {\bf 58}, 51 (2008)  [arXiv:0801.3867 [hep-ph]].

\bibitem{Drenska:2010kg}
  N.~Drenska, R.~Faccini, F.~Piccinini, A.~Polosa, F.~Renga and C.~Sabelli,
  Riv.\ Nuovo Cim.\  {\bf 033}, 633 (2010)  [arXiv:1006.2741 [hep-ph]].

\bibitem{Bodwin:2013nua}
  G.~T.~Bodwin, E.~Braaten, E.~Eichten, S.~L.~Olsen, T.~K.~Pedlar and J.~Russ,
  arXiv:1307.7425.

\bibitem{Choi:2003ue}
  S.~K.~Choi {\it et al.}  [Belle Collaboration],
  Phys.\ Rev.\ Lett.\  {\bf 91}, 262001 (2003)
  [hep-ex/0309032].


\bibitem{Aubert:2004ns}
  B.~Aubert {\it et al.}  [BaBar Collaboration],
  Phys.\ Rev.\ D {\bf 71}, 071103 (2005)
  [hep-ex/0406022].


\bibitem{Abazov:2004kp}
  V.~M.~Abazov {\it et al.}  [D0 Collaboration],
  Phys.\ Rev.\ Lett.\  {\bf 93}, 162002 (2004)
  [hep-ex/0405004].


\bibitem{Aaltonen:2009vj}
  T.~Aaltonen {\it et al.}  [CDF Collaboration],
  Phys.\ Rev.\ Lett.\  {\bf 103}, 152001 (2009)
  [arXiv:0906.5218 [hep-ex]].


\bibitem{Chatrchyan:2013cld}
  S.~Chatrchyan {\it et al.}  [CMS Collaboration],
  JHEP {\bf 1304}, 154 (2013)
  [arXiv:1302.3968 [hep-ex]].


\bibitem{Aaij:2013zoa}
  RAaij {\it et al.}  [LHCb Collaboration],
  Phys.\ Rev.\ Lett.\  {\bf 110}, no. 22, 222001 (2013)
  [arXiv:1302.6269 [hep-ex]].


\bibitem{Beringer:1900zz}
  J.~Beringer {\it et al.}  [Particle Data Group Collaboration],
  Phys.\ Rev.\ D {\bf 86}, 010001 (2012).


\bibitem{TheBABAR:2013dja}
  J. P. Lees {\it et al.}  [ The BABAR Collaboration],
  Phys.\ Rev.\ D {\bf 88}, 071104 (2013)
  [arXiv:1308.1151 [hep-ex]].


\bibitem{Tornqvist:2004qy}
  N.~A.~Tornqvist,
  Phys.\ Lett.\ B {\bf 590}, 209 (2004)
  [hep-ph/0402237].


\bibitem{Hanhart:2007yq}
  C.~Hanhart, Y.~.S.~Kalashnikova, A.~E.~Kudryavtsev and A.~V.~Nefediev,
  Phys.\ Rev.\ D {\bf 76}, 034007 (2007)
  [arXiv:0704.0605 [hep-ph]].


\bibitem{Hou:2006it}
  W.~-S.~Hou,
  Phys.\ Rev.\ D {\bf 74}, 017504 (2006)
  [hep-ph/0606016].


\bibitem{Aushev:2010bq}
  T.~Aushev, W.~Bartel, A.~Bondar, J.~Brodzicka, T.~E.~Browder, P.~Chang, Y.~Chao and K.~F.~Chen {\it et al.},
  arXiv:1002.5012 [hep-ex].



\bibitem{GMW}
  F.~-K.~Guo, U.~-G.~Mei{\ss}ner and W.~Wang,
  arXiv:1402.6236 [hep-ph].


\bibitem{Bignamini:2009sk}
  C.~Bignamini, B.~Grinstein, F.~Piccinini, A.~D.~Polosa and C.~Sabelli,
  Phys.\ Rev.\ Lett.\  {\bf 103}, 162001 (2009)
  [arXiv:0906.0882 [hep-ph]].

\bibitem{Esposito:2013ada}
  A.~Esposito, F.~Piccinini, A.~Pilloni and A.~Polosa,
  arXiv:1305.0527 [hep-ph].


\bibitem{Artoisenet:2009wk}
  P.~Artoisenet and E.~Braaten,
  Phys.\ Rev.\ D {\bf 81}, 114018 (2010)
  [arXiv:0911.2016 [hep-ph]].


\bibitem{Artoisenet:2010uu}
  P.~Artoisenet and E.~Braaten,
  Phys.\ Rev.\ D {\bf 83}, 014019 (2011)
  [arXiv:1007.2868 [hep-ph]].





\bibitem{Ali:2011qi}
  A.~Ali and W.~Wang,
  Phys.\ Rev.\ Lett.\  {\bf 106}, 192001 (2011)
  [arXiv:1103.4587 [hep-ph]].


\bibitem{Ali:2013xba}
  A.~Ali, C.~Hambrock and W.~Wang,
  Phys.\  Rev.\  D 88, {\bf 054026} (2013)
  [arXiv:1306.4470 [hep-ph]].


\bibitem{Guo:2013ufa}
  F.~-K.~Guo, U.~-G.~Mei{\ss}ner and W.~Wang,
  arXiv:1308.0193 [hep-ph], to appear in Communications in Theoretical Physics. 


\bibitem{Chatrchyan:2013mea}
  S.~Chatrchyan {\it et al.}  [CMS Collaboration],
  Phys.\ Lett.\ B {\bf 727}, 57 (2013)
  [arXiv:1309.0250 [hep-ex]].


\bibitem{Lipkin:1986bi}
  H.~J.~Lipkin,
  Nucl.\ Phys.\ B {\bf 291}, 720 (1987).


\bibitem{Lipkin:1988tg}
  H.~J.~Lipkin and S.~F.~Tuan,
  Phys.\ Lett.\ B {\bf 206}, 349 (1988).


\bibitem{Moxhay:1988ri}
  P.~Moxhay,
  Phys.\ Rev.\ D {\bf 39}, 3497 (1989).


\bibitem{Wang:2013cya}
  Q.~Wang, C.~Hanhart and Q.~Zhao,
  Phys.\ Rev.\ Lett.\  {\bf 111}, 132003 (2013)
  [arXiv:1303.6355 [hep-ph]].


\bibitem{Liu:2013vfa}
  X.~-H.~Liu and G.~Li,
  Phys.\ Rev.\ D {\bf 88}, 014013 (2013)
  [arXiv:1306.1384 [hep-ph]].


\bibitem{Guo:2013zbw}
  F.~-K.~Guo, C.~Hanhart, U.~-G.~Mei{\ss}ner, Q.~Wang and Q.~Zhao,
  Phys.\ Lett.\ B {\bf 725}, 127 (2013)
  [arXiv:1306.3096 [hep-ph]].


\bibitem{Wang:2013hga}
  Q.~Wang, C.~Hanhart and Q.~Zhao,
  Phys.\ Lett.\ B {\bf 725}, no. 1-3, 106 (2013)
  [arXiv:1305.1997 [hep-ph]].


\bibitem{Cleven:2013sq}
  M.~Cleven, Q.~Wang, F.~-K.~Guo, C.~Hanhart, U.~-G.~Mei{\ss}ner and Q.~Zhao,
  Phys.\ Rev.\ D {\bf 87}, no. 7, 074006 (2013)
  [arXiv:1301.6461 [hep-ph]].


\bibitem{Chen:2011pv}
  D.~-Y.~Chen and X.~Liu,
  Phys.\ Rev.\ D {\bf 84}, 094003 (2011)
  [arXiv:1106.3798 [hep-ph]].


\bibitem{Li:2012as}
  G.~Li, F.~-l.~Shao, C.~-W.~Zhao and Q.~Zhao,
  Phys.\ Rev.\ D {\bf 87}, no. 3, 034020 (2013)
  [arXiv:1212.3784 [hep-ph]].


\bibitem{Li:2013yla}
  G.~Li and X.~-H.~Liu,
  Phys.\ Rev.\ D {\bf 88},  094008 (2013)
  [arXiv:1307.2622 [hep-ph]].

\bibitem{Voloshin:2013ez}
  M.~B.~Voloshin,
  Phys.\ Rev.\ D {\bf 87}, no. 7, 074011 (2013)  [arXiv:1301.5068 [hep-ph]].

\bibitem{Voloshin:2011qa}
  M.~B.~Voloshin,
  Phys.\ Rev.\ D {\bf 84}, 031502 (2011)  [arXiv:1105.5829 [hep-ph]].

\bibitem{Bondar:2011ev}
  A.~E.~Bondar, A.~Garmash, A.~I.~Milstein, R.~Mizuk and M.~B.~Voloshin,
  Phys.\ Rev.\ D {\bf 84}, 054010 (2011)  [arXiv:1105.4473 [hep-ph]].


\bibitem{Guo:2009wr}
  F.~-K.~Guo, C.~Hanhart and U.~-G.~Mei{\ss}ner,
  Phys.\ Rev.\ Lett.\  {\bf 103}, 082003 (2009)
  [Erratum-ibid.\  {\bf 104}, 109901 (2010)]
  [arXiv:0907.0521 [hep-ph]].


\bibitem{oai:arXiv.org:1002.2712}
  F.~-K.~Guo, C.~Hanhart, G.~Li, U.~-G.~Meissner and Q.~Zhao,
  Phys.\ Rev.\ D {\bf 82}, 034025 (2010)
  [arXiv:1002.2712 [hep-ph]].

\bibitem{Guo:2010ak}
  F.~-K.~Guo, C.~Hanhart, G.~Li, U.~-G.~Mei{\ss}ner and Q.~Zhao,
  Phys.\ Rev.\ D {\bf 83}, 034013 (2011)
  [arXiv:1008.3632 [hep-ph]].

\bibitem{Chen:2011pu}
  D.~-Y.~Chen, X.~Liu and T.~Matsuki,
  Phys.\ Rev.\ D {\bf 84}, 074032 (2011)
  [arXiv:1108.4458 [hep-ph]].

\bibitem{Chen:2012yr}
  D.~-Y.~Chen, X.~Liu and T.~Matsuki,
  arXiv:1208.2411 [hep-ph].

\bibitem{Chen:2013coa}
  D.~-Y.~Chen, X.~Liu and T.~Matsuki,
  Phys.\ Rev.\ D {\bf 88}, 036008 (2013)
  [arXiv:1304.5845 [hep-ph]].

\bibitem{Chen:2013bha}
  D.~-Y.~Chen, X.~Liu and T.~Matsuki,
  Phys.\ Rev.\ D {\bf 88}, 014034 (2013)
  [arXiv:1306.2080 [hep-ph]].


\bibitem{Du:1998ss}
  D.~-S.~Du, X.~-Q.~Li, Z.~-T.~Wei and B.~-S.~Zou,
  Eur.\ Phys.\ J.\ A {\bf 4}, 91 (1999)
  [hep-ph/9805260].

\bibitem{Chen:2000ih}
  C.~-H.~Chen and H.~-n.~Li,
  Phys.\ Rev.\ D {\bf 63}, 014003 (2001)
  [hep-ph/0006351].
\bibitem{Liu:2007qs}
  X.~Liu and X.~-Q.~Li,
  Phys.\ Rev.\ D {\bf 77}, 096010 (2008)
  [arXiv:0707.0919 [hep-ph]].

\bibitem{Colangelo:2002mj}
  P.~Colangelo, F.~De Fazio and T.~N.~Pham,
  Phys.\ Lett.\ B {\bf 542}, 71 (2002)
  [hep-ph/0207061].

\bibitem{Cheng:2004ru}
  H.~-Y.~Cheng, C.~-K.~Chua and A.~Soni,
  Phys.\ Rev.\ D {\bf 71}, 014030 (2005)
  [hep-ph/0409317].

\bibitem{Lu:2005mx}
  C.~-D.~Lu, Y.~-L.~Shen and W.~Wang,
  Phys.\ Rev.\ D {\bf 73}, 034005 (2006)
  [hep-ph/0511255].

\bibitem{Liu:2008tv}
  X.~Liu, Z.~-T.~Wei and X.~-Q.~Li,
  Eur.\ Phys.\ J.\ C {\bf 59}, 683 (2009)
  [arXiv:0805.2804 [hep-ph]].

\bibitem{Colangelo:2003sa}
  P.~Colangelo, F.~De Fazio and T.~N.~Pham,
  Phys.\ Rev.\ D {\bf 69}, 054023 (2004)
  [hep-ph/0310084].


\bibitem{Casalbuoni:1996pg}
  R.~Casalbuoni, A.~Deandrea, N.~Di Bartolomeo, R.~Gatto, F.~Feruglio and G.~Nardulli,
  Phys.\ Rept.\  {\bf 281}, 145 (1997)
  [hep-ph/9605342].


\bibitem{Weinberg:1965zz}
  S.~Weinberg,
  Phys.\ Rev.\  {\bf 137}, B672 (1965).


\bibitem{Baru:2003qq}
  V.~Baru, J.~Haidenbauer, C.~Hanhart, Y.~Kalashnikova and A.~E.~Kudryavtsev,
  Phys.\ Lett.\ B {\bf 586}, 53 (2004)
  [hep-ph/0308129].


\bibitem{Hu:2005gf}
  J.~Hu and T.~Mehen,
  Phys.\ Rev.\ D {\bf 73}, 054003 (2006)
  [hep-ph/0511321].


\bibitem{Amundson:1992yp}
  J.~F.~Amundson, C.~G.~Boyd, E.~E.~Jenkins, M.~E.~Luke, A.~V.~Manohar, J.~L.~Rosner, M.~J.~Savage and M.~B.~Wise,
  Phys.\ Lett.\ B {\bf 296}, 415 (1992)
  [hep-ph/9209241].


\bibitem{Li:1996yn}
  X.~-Q.~Li, D.~V.~Bugg and B.~-S.~Zou,
  Phys.\ Rev.\ D {\bf 55}, 1421 (1997).


\bibitem{Locher:1993cc}
  M.~P.~Locher, Y.~Lu and B.~S.~Zou,
  Z.\ Phys.\ A {\bf 347}, 281 (1994)
  [nucl-th/9311021].


\bibitem{Li:1996cj}
  X.~-Q.~Li and B.~-S.~Zou,
  Phys.\ Lett.\ B {\bf 399}, 297 (1997)
  [hep-ph/9611223].


\bibitem{Zhao:2013jza}
  C.~-W.~Zhao, G.~Li, X.~-H.~Liu and F.~-L.~Shao,
  Eur.\ Phys.\ J.\ C {\bf 73}, 2482 (2013).




\bibitem{Ali:2009pi}
  A.~Ali, C.~Hambrock, I.~Ahmed and M.~J.~Aslam,
  Phys.\ Lett.\ B {\bf 684}, 28 (2010)
  [arXiv:0911.2787 [hep-ph]].


\bibitem{Tornqvist:1993ng}
  N.~A.~Tornqvist,
  Z.\ Phys.\ C {\bf 61}, 525 (1994)
  [hep-ph/9310247].


\bibitem{Guo:2013sya}
  F.~-K.~Guo, C.~Hidalgo-Duque, J.~Nieves and M.~P.~Valderrama,
  Phys.\ Rev.\ D {\bf 88}, 054007 (2013)
  [arXiv:1303.6608 [hep-ph]].


\bibitem{Karliner:2013dqa}
  M.~Karliner and S.~Nussinov,
  JHEP {\bf 1307}, 153 (2013)
  [arXiv:1304.0345 [hep-ph]].




 \end{thebibliography}
\end{document}